\documentclass[twocolumn]{aastex631}
\graphicspath{{./}{figures/}}

\usepackage[utf8]{inputenc}
\usepackage[T1]{fontenc}

\begin{document}

\title{Modulation of the Blazhko Cycle in LS Her}
\correspondingauthor{Ronald Wilhelm}
\email{ron.wilhelm@uky.edu}

\author[0000-0002-4792-7722]{Ronald Wilhelm}
\affiliation{Department of Physics \& Astronomy, University of Kentucky, 177 Chem.-Phys.~Building, 505 Rose Street, Lexington, KY 40506, USA}

\author[0000-0002-6307-992X]{Kenneth Carrell}
\affiliation{Physics \& Geosciences, Angelo State University, 2601 W.~Avenue N, San Angelo, TX 76909, USA}

\author{Hannah H. Means}
\affiliation{Physics \& Astronomy, Bowling Green State University, Bowling Green, OH 43403, USA}

\author[0000-0003-3184-5228]{Adam Popowicz}
\affiliation{Department  of Electronics, Electrical Engineering, \& Microelectronics, Silesian University of Technology, Akademicka 16, 44-100 Gliwice, Poland}
%% EMAIL: (apopowicz@polsl.pl)

\author[0000-0003-4647-7114]{Krzysztof Bernacki}
\affiliation{Department  of Electronics, Electrical Engineering, \& Microelectronics, Silesian University of Technology, Akademicka 16, 44-100 Gliwice, Poland}
%% EMAIL: (kbernacki@polsl.pl )

\author[0000-0002-3147-2348]{Mariusz Fr\k{a}ckiewicz}
\affiliation{Department of Data Science \& Engineering, Silesian University of Technology, Akademicka 16, 44-100 Gliwice, Poland}
%% EMAIL: (mfrackiewicz@polsl.pl )

\author[0000-0003-1501-2154]{Marek Szczepa\'{n}ski}
\affiliation{Department of Data Science \& Engineering, Silesian University of Technology, Akademicka 16, 44-100 Gliwice, Poland}
%% EMAIL: (mszczepanski@polsl.pl)

\author[0000-0001-5516-9733]{Adam Dustor}
\affiliation{Department of Telecommunications \& Teleinformatics, Silesian University of Technology, Akademicka 16, 44-100 Gliwice, Poland}
%% EMAIL: (adustor@polsl.pl)

\author{Jaylon Lockett}
\affiliation{Physics \& Geosciences, Angelo State University, 2601 W.~Avenue N, San Angelo, TX 76909, USA}

\author{Jill Taylor}
\affiliation{Physics \& Geosciences, Angelo State University, 2601 W.~Avenue N, San Angelo, TX 76909, USA}

\author{Stasha Youngquist}
\affiliation{Physics \& Geosciences, Angelo State University, 2601 W.~Avenue N, San Angelo, TX 76909, USA}

\date{February 2023}

\begin{abstract}
    We present analysis of the RR Lyrae star, LS Her and confirm the previously reported modulation to its Blazhko cycles.  We
performed Fourier analysis on two sectors (Sector 24 \& 25) of data from the Transiting Exoplanet Survey Satellite
(TESS) spanning 53 days.  We find LS Her to have a primary pulsation period of 0.2308 d and a
Blazhko period of 12.7 d in keeping with previously reported results.  We also identified side-band
frequencies around the Blazhko multiplets suggesting the Blazhko cycle is modulated on a time scale of
112 days.  Analysis of the Blazhko effect using the TESS data clearly shows a changing amplitude and phase
throughout the four Blazhko cycles.  We compared our modeled results, which were based on our TESS frequency analysis, to TESS data (Sector 51) taken $\sim$700 days later
and found our modulation model was not a good representation of the data. We then coupled our TESS analysis with the modulation frequency results
from \citet{wils08} and found excellent agreement with the Sector 51
data.  To further test this result we obtained ground-based, V-magnitude observations of LS Her in the summer of
2022. This data also showed excellent agreement with our coupled modulation model.  We have verified that LS Her is
a Blazhko star with a modulated Blazhko period of 109 days, stability over the 862 days of observations, and possible stability lasting over 15 years.  We discuss the ramifications of the modulation for other Blazhko
stars that show Blazhko effect changes over time.
\end{abstract}

\section{Introduction}\label{sec:intro}
The Blazhko effect (BE) in the radial pulsating RR Lyrae (RRL) stars was first recognized over a century ago 
by Sergey Blazhko (\citeyear{blazhko}).  The effect can often be seen in lightcurve data as a cyclic variation 
in the amplitude and/or phase of the radial pulsation.  It can also be detected in the frequency spectrum as multiplet 
peaks on either side of the primary pulsation frequency peak as well as the lower order harmonics.  The frequency 
separation between these side-bands and the primary frequency is the Blazhko frequency, which can have periods as small 
as a few days or as long as several years.  The beating between the primary and Blazhko frequencies results in 
a modulation of the lightcurve, which causes cyclic changes to the pulsation amplitude and/or phase.

The cause of the BE remains unknown, although the non-linear mode resonance model \citep{buchler11,kollath18} 
is seen as a current, favored hypothesis.  There remains no hydrodynamic simulation that can 
successfully model this effect.  

In recent years, instability in the BE has been identified in some Blazhko stars, which result in changes to 
the Blazhko period (BP) and amplitude over time.  For these stars, the BE modulates the individual 
radial pulsations, but the BE itself is not always stable across many Blazhko cycles.  It is currently unclear 
whether the BE changes are irregular or periodically modulated by an even longer modulation period (MP).  Understanding 
these changes to the BE is important to help constrain both internal and external effects that may play a role 
in the BE phenomenon. 

In the past, there have been a number of Blazhko stars found to change their Blazhko period and amplitude 
over time spans much longer than their BPs. \citet{detre73} found that the star, RR Lyrae, has a 
changing Blazhko period and amplitude that seems to occur on a 4 year 
cycle.  Further research by \citet{kolenberg06} 
also found changes in RR Lyrae's BP, suggesting that the BE is possibly 
being modulated on this 4 year cycle.  However, \citet{leborgne14} 
found the 4-year modulation is not consistent with changes seen from Kepler data.  The authors find a decline in the BE over 5 years but not the 4-year modulation. 

The RRc variable, TV Boo, was found to have modulated Blazhko changes with a period of 21.5 days \citep{skarka13} 
while XZ Cyg was found to have Blazhko period and amplitude changes on the order of a decade \citep{lacluyze04}. 
There are also clear Blazhko changes found in RRL stars in the globular cluster M3 \citep{jurcsik12}, 
where previously known Blazhko stars have become non-Blazhko variables. 

\citet{wils08} discovered a modulated BE in a first over-tone pulsating, RRc star, LS Herculis.
The authors found LS Her to be a very short period variable star (P = 0.232 d) with a short 
BP = 12.75 days.  \citeauthor{wils08}~observed this star for 63 nights over a span of six months 
in 2007.  Their analysis found equally spaced 
side-bands around the Blazhko side-bands, indicating that the BE itself was being modulated.  This modulation 
period was found to be MP = 109 d.  

Note that many papers exploring the BE use the term ``modulated Blazhko effect'' to clarify that 
the BE is modulating the primary pulsation.  For the rest of this paper we will refer to modulation in terms 
of changes to the BE itself.  

In Section \ref{sec:tess} we present the frequency analysis of two Transiting Exoplanet Survey Satellite 
\citep[TESS,][]{tess} sectors for LS Her.  In Section 
\ref{sec:wils} we construct a model lightcurve using the modulation frequencies from \citeauthor{wils08}~and 
compare this model to the TESS data from Sector 51.  Section \ref{sec:grou} compares our 
recent ground-based observations from the Silesian University Technology Observatories
(SUTO) to that of our modeled lightcurve and Section \ref{sec:disc} is a discussion on
ramifications of our findings. 

\section{TESS data analysis}\label{sec:tess}

\begin{figure}[htb]
\plotone{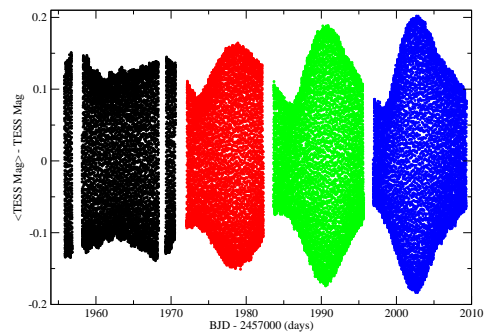}
\caption{TESS data from Sectors 24 \& 25.  The BE is clearly seen to change from one Blazhko cycle to the next with the amplitude of the BE growing with each successive cycle. The color coding indicates each TESS orbit.\label{fig:lightcurve}}
\end{figure}

LS Her has been observed three times with the TESS 2-minute integration program. 
TESS Sectors 24 (April 16, 2020 to May 13, 2020) and 25 (May 13, 2020 to June 8, 2020) were observed back-to-back giving 
nearly continuous observations for a time span of $\sim$53 days. It was later observed in Sector 51 (April 22, 2022 to 
May 18, 2022) as part of the approved Guest Investigator Program \#G04184 (submitted by Carrell). All the TESS data used in this paper can be found in MAST: \dataset[10.17909/svrw-cn60]{http://dx.doi.org/10.17909/svrw-cn60}.
The lightcurve for Sectors 24 \& 25 can be seen in Figure \ref{fig:lightcurve}, where the time axis is in TESS Barycentric Julian Date (BJD). 
There are several clearly recognizable characteristics of this data.  First, cyclic changes in the pulsation amplitude as a 
function of time validates the classification of LS Her as a Blazhko variable.  Second, the Blazhko amplitude appears 
to grow as a function of time, corroborating the findings of \citet{wils08}, that this star experiences a BE which itself 
is being modulated.  Finally, from the first orbit data (black) of Sector 24 it appears that during some portions of the 
modulation, the Blazhko amplitude is severely damped,
with the difference between B$_{\textrm{max}}$ and B$_{\textrm{min}}$ becoming only a small fraction of what is seen in later Blazhko cycles.

\begin{figure*}[htb]
  \centering
  \begin{minipage}{0.4\textwidth}
    \centering
    \includegraphics[width=1.0\textwidth]{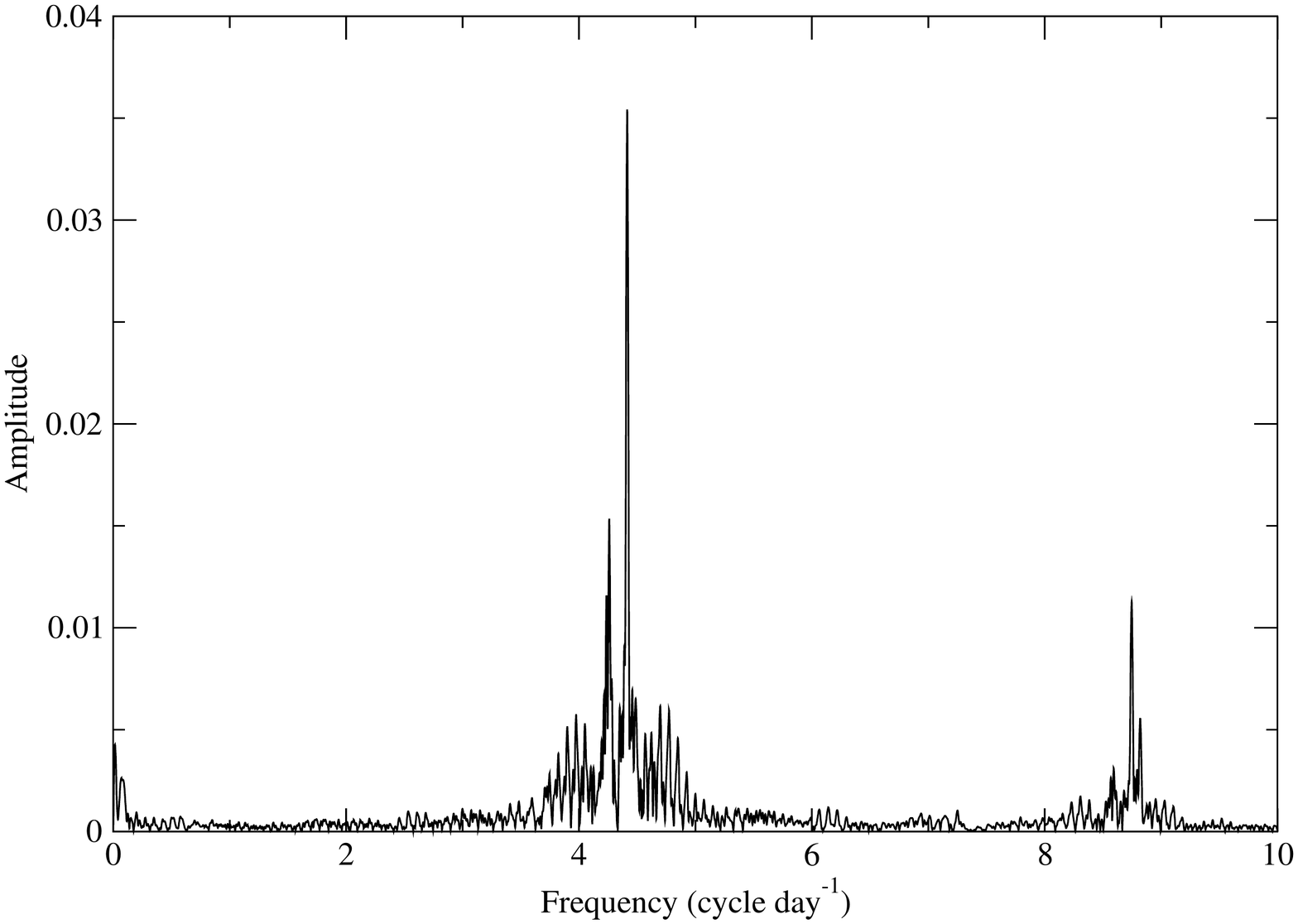}
    \label{fig:FW1}
  \end{minipage}
    \begin{minipage}{0.4\textwidth}
    \centering
    \includegraphics[width=1.0\textwidth]{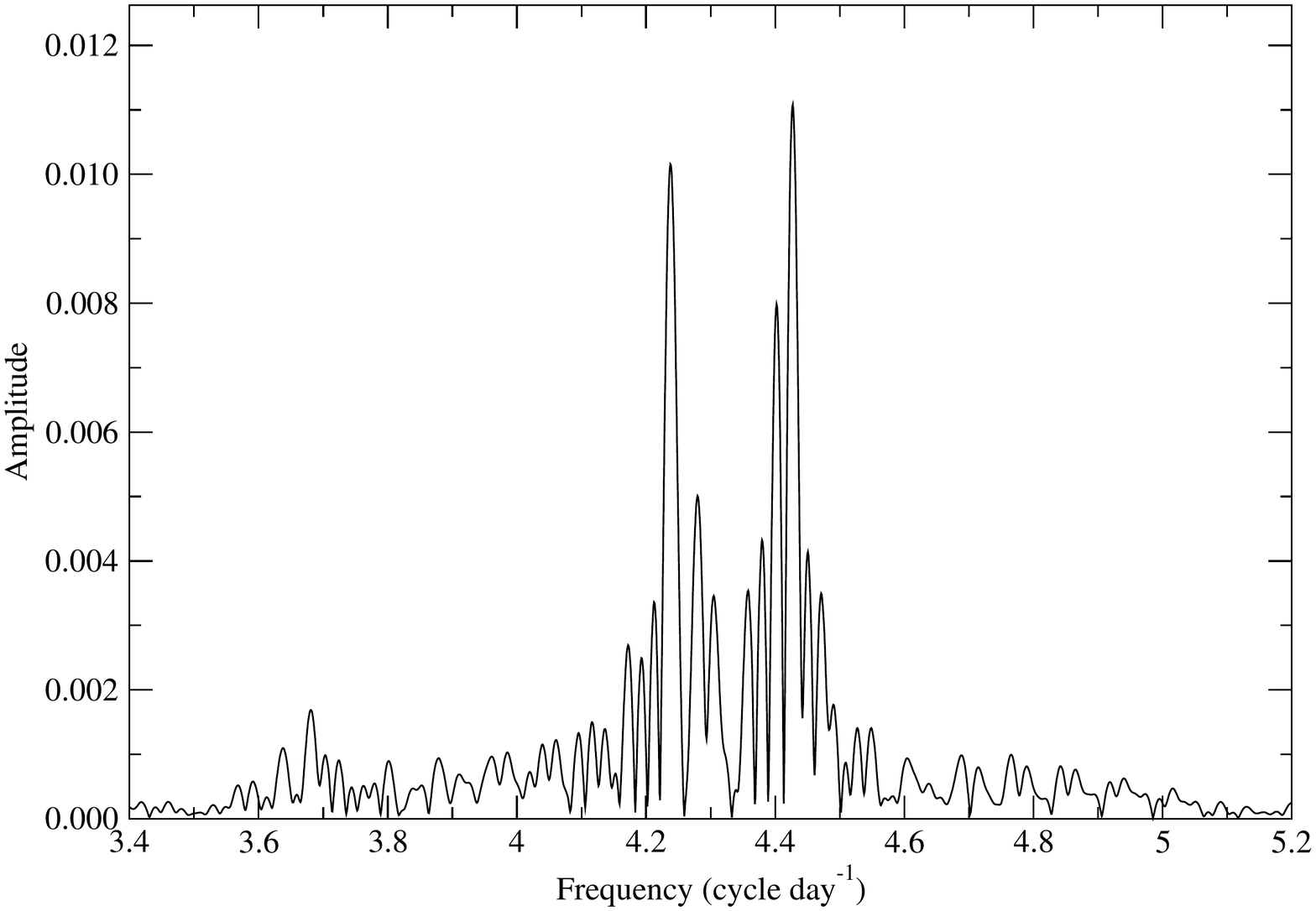}
    \label{fig:F3b}
      \end{minipage}
  \caption{The Fourier spectrum from TESS data after pre-whittening of the primary frequency and 
  higher harmonics. The Blazhko side-bands are promenient near the primary and 2nd harmonic peaks. 
  The right panel shows modulation peaks after pre-whittening the Blazhko side-bands\label{fig:fourier}}
\end{figure*}

\begin{figure}[htb]
\plotone{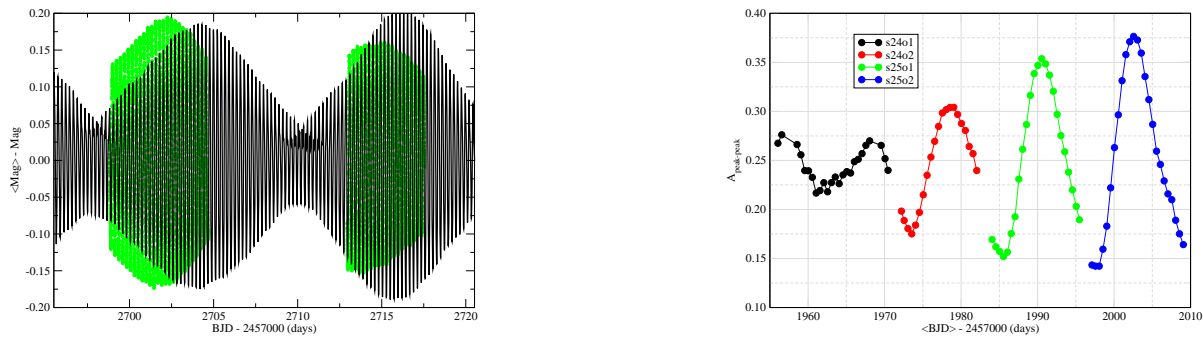}
\caption{Comparison of Sector 51 data (green) and the modeled data from Sectors 24 \& 25 (black). There is a time misalignment and a less than satisfactory amplitude match with the second orbit amplitude being much smaller than the model data.\label{fig:sec51misalign}}
\end{figure}

\begin{figure}[htb]
\plotone{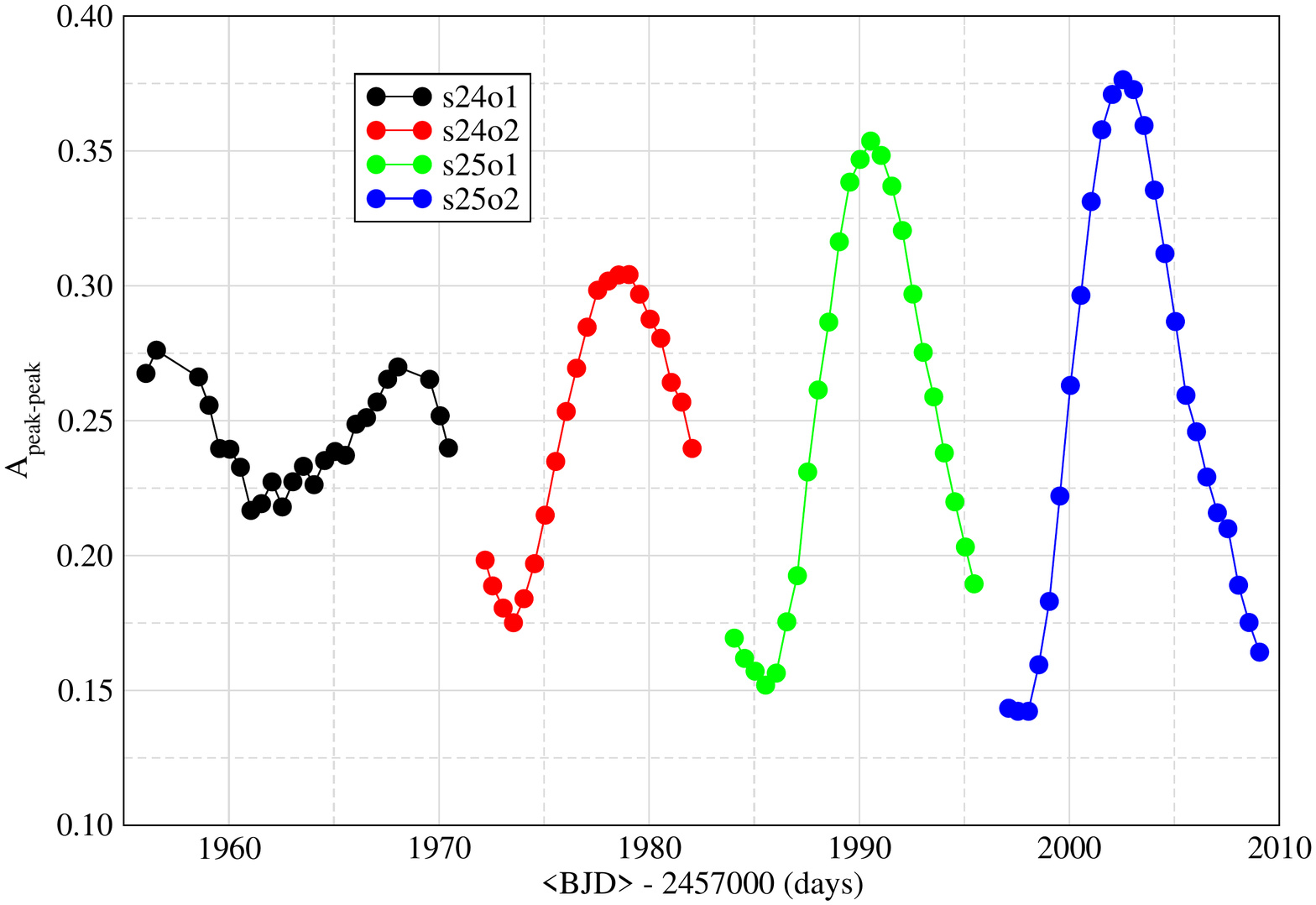}
\caption{The peak-to-peak amplitude variations as a function of BJD for Sectors 24 \& 25.  The color coding is split by TESS orbits. The amplitude was measured from the fitting of the 0.5 day segments described in the text.\label{fig:peak2peak}}
\end{figure}

\begin{figure*}[htb]
  \centering
  \includegraphics[width=1.0\textwidth]{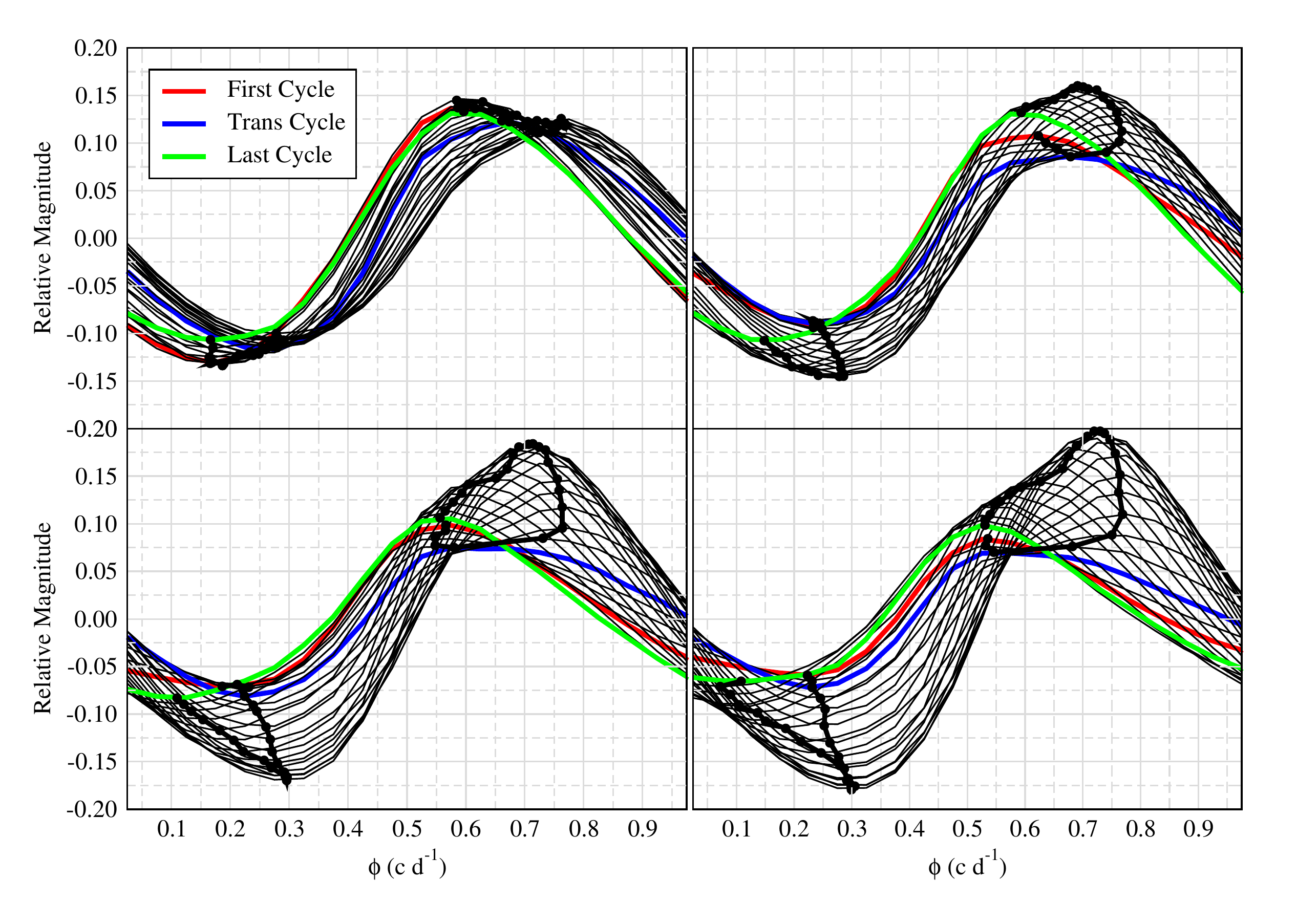}
  \caption{Comparison of the sub-groups for each TESS orbit.  The top figures are for Sector 24 and bottom Sector 25, with orbit 1 on the left for both sectors.  The red line is the first sub-group while the green line marks the final for each orbit.  The blue line indicates when the maximum shifts from the hump near $\phi$ = 0.55 to the maximum near $\phi$ = 0.75.  The heavy black line marks the maximum (minimum) for each sub-group.\label{fig:subgroups}}
\end{figure*}

\begin{figure}[htb]
\plotone{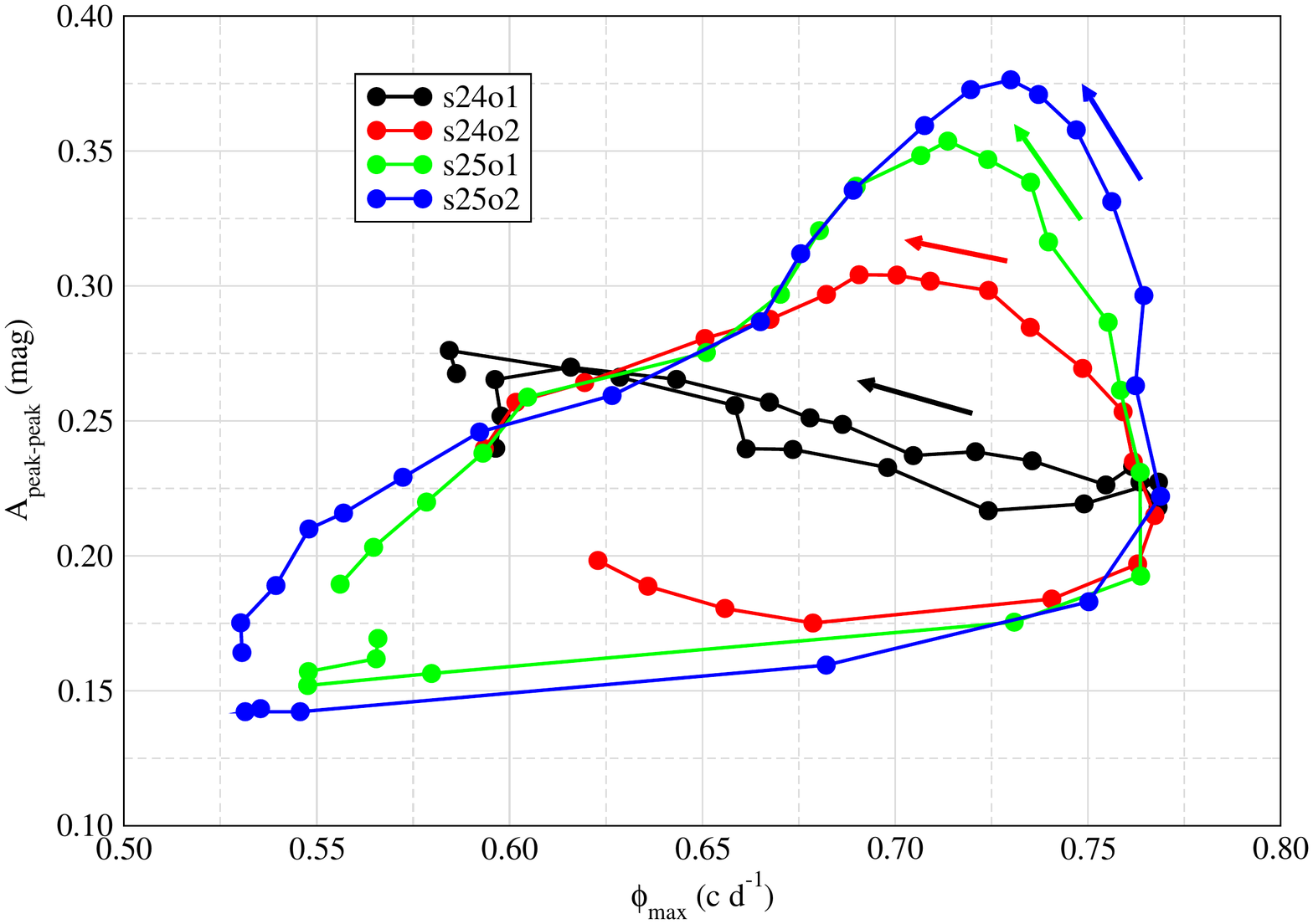}
\caption{A phase-amplitude diagram showing the changes from one TESS orbit to the next. The arrows show the direction of the temporal changes which are counter-clockwise in all cases. \label{fig:phaseamplitude}}
\end{figure}

We used Period04 \citep{period04} to analyze the Sector 24 \& 25 data. 
In order to place all the data on a consistent scale, we normalized on a by-orbit 
basis to the overall flux average. After converting the normalized flux values to 
magnitudes, we subtract the mean magnitude to put the average magnitude at zero.
Using Period04, we find the primary pulsation 
frequency to be virtually the same as \citeauthor{wils08}~with F$_0$ = 4.3326 c d$^{-1}$. 
The left panel in Figure \ref{fig:fourier} shows the Fourier spectrum after pre-whitening the 
primary and significant harmonics.  Side-bands are clearly present around the 
primary frequency and the 2nd harmonic.  The right panel shows the frequencies 
remaining after pre-whittening the primary Blazhko frequencies.  These peaks 
correspond to the modulation peaks found by \citeauthor{wils08} Table \ref{tab:results} 
lists the frequencies, amplitudes and phases we found for all peaks with S/N $>$ 8. 
Although there are still some frequency peaks remaining that would be considered significant, 
beyond the lines in the table, Period04 was not able to converge on a solution.  We note that the F$_\textrm{u}$ frequencies are significant but of unknown origin.

Along with our agreement for the primary pulsation frequency, the next three harmonics also agree with \citeauthor{wils08}~to within the uncertainty. \citeauthor{wils08}~found Blazhko triplets around the primary and 2nd harmonic and one side-band around the 3rd harmonic.  From these, they find a BP of 12.75 $\pm$ 0.02 d.  We identified triplets around the first four harmonics (F$_\textrm{B}$) which gave us a BP of 12.69 $\pm$ 0.30 d, very close to their determination.   We were unable to clearly identify the modulation triplets from the TESS data but we were able to identify several modulation peaks (F$_\textrm{m}$) that are very close to the \citeauthor{wils08}~values.  Our determination of the MP was 112.3 $\pm$ 5.6 d, compared to their value of 109 $\pm$ 4 d.  We expect that our much shorter baseline of 53 days results in blending of the frequency peaks making the TESS data results more uncertain, particularly since our data covers less than 50\% of the MP.

\begin{table}[h!]
  \begin{center}
    \caption{Fourier Analysis Results.}
    \label{tab:results}
    \begin{tabular}{|c|D|D|D|}
    \hline
     \textbf{Identification} & \multicolumn{2}{c|}{\textbf{Frequency}} & \multicolumn{2}{c|}{\textbf{Amplitude}} & \multicolumn{2}{c|}{$\phi$}\\
      & \multicolumn{2}{c|}{(c d$^{-1}$)} & \multicolumn{2}{c|}{(mmag)} & \multicolumn{2}{c|}{} \\
      \hline
\decimals
F$_0$      &  4.3326	& 116.11 & 0.435 \\
2F$_0$     &  8.6653	&  12.57 & 0.119 \\
3F$_0$     & 12.9980	&   4.55 & 0.235 \\
4F$_0$     & 17.3306	&   2.49 & 0.931 \\
5F$_0$     & 21.6632	&   1.27 & 0.913 \\
\hline
F$_0$ + F$_{\textrm{B}}$                    & 4.4126 & 40.53 & 0.259 \\ 
F$_0$ - F$_{\textrm{B}}$                    & 4.2544 & 26.57 & 0.891 \\
F$_0$ + F$_{\textrm{B}}$ + F$_{\textrm{m}}$ & 4.4198 & 21.01 & 0.763 \\
F$_0$ - F$_{\textrm{B}}$ - F$_{\textrm{m}}$ & 4.2446 & 20.11 & 0.676 \\
F$_0$ - F$_{\textrm{B}}$ + F$_{\textrm{m}}$ & 4.2645 &  6.78 & 0.127 \\
\hline
2F$_0$ + F$_{\textrm{B}}$                     & 8.7452 & 9.54 & 0.532 \\
2F$_0$ - F$_{\textrm{B}}$                     & 8.5910 & 3.91 & 0.353 \\
2F$_0$ + F$_{\textrm{B}}$ + F$_{\textrm{m}}$  & 8.7537 & 4.59 & 0.409 \\
2F$_0$ + 2F$_{\textrm{B}}$                    & 8.8223 & 4.06 & 0.097 \\
2F$_0$ - F$_{\textrm{B}}$ - 2F$_{\textrm{m}}$ & 8.5716 & 2.30 & 0.193 \\
\hline
3F$_0$ + F$_{\textrm{B}}$  & 13.0794 & 2.34 & 0.736 \\
3F$_0$ - F$_{\textrm{B}}$  & 12.9007 & 0.96 & 0.922 \\
3F$_0$ - 2F$_{\textrm{B}}$ & 12.8423 & 1.36 & 0.342 \\
3F$_0$ + 2F$_{\textrm{B}}$ & 13.1545 & 0.81 & 0.131 \\
\hline
4F$_0$ + F$_{\textrm{B}}$  & 17.4115 & 1.34 & 0.439 \\
4F$_0$ - 2F$_{\textrm{B}}$ & 17.1741 & 1.55 & 0.518 \\
\hline
F$_{\textrm{B}}$  & 0.0830 & 4.80 & 0.527 \\
2F$_{\textrm{B}}$ & 0.1617 & 1.21 & 0.942 \\
\hline
F$_{\textrm{1u}}$   & 0.1504 & 1.74 & 0.692 \\ 
F$_{\textrm{2u}}$   & 3.6780 & 1.40 & 0.647 \\
F$_{\textrm{2u}^\prime}$  & 3.6389 & 0.97 & 0.450 \\
2F$_{\textrm{2u}^\prime}$ & 7.2517 & 0.87 & 0.879 \\
\hline
    \end{tabular}
  \end{center}
\end{table}

To test our results, we made use of the more recent TESS data from Sector 51. 
This data was taken approximately 700 days 
after the end of our modeled TESS data.  A comparison of the extrapolated model 
and the Sector 51 data can be seen in Figure \ref{fig:sec51misalign}. The 
comparison is less than satisfactory with the Sector 51 data being shifted 
by several days compared to the model lightcurve.  Also, the modulation 
envelope shape is very different from the TESS data, particularly for orbit 
2, where the data shows a declining Blazhko amplitude instead of the increase 
seen in the model.  Under the assumption that the MP has been 
constant over the 700 days, we conclude that our model is not able to fully capture 
the modulation effect. This is likely caused by the blending of some frequency peaks due to the relatively short baseline of observations.  In the upcoming sections we will show that the 
\citeauthor{wils08}~modulation measurements are superior to the measurements of our
TESS data sample.

To further analyze the LS Her lightcurve, we used the results from Sectors 
24 \& 25 to produce a phased lightcurve.  The final phased 
data was then divided into time segments of 0.5 d.  Given that the pulsation 
period of LS Her 
is P $\sim$0.231 d, each sub-group contained just over two full pulsation 
cycles.  We used the Period04 average data binning feature 
to fit a curve to the phase data for each of these sub-groups.  We set the binning to 20 
($\Delta\phi$ = 0.05), which was able to represent the light curve well and still provide
$\sim$18 data points per bin over which to average. Although LS Her is a 
Blazhko star with changing amplitude, the combining of consecutive pulsation 
cycles resulted in an insignificant scatter of a few millimags to the fit.  We 
determined the value of maximum and minimum from the fit of each sub-group and 
produced a corresponding peak-to-peak amplitude.  The result can be seen in 
Figure \ref{fig:peak2peak} where the data are color coded by sector and 
orbit.  The BP is very similar to the TESS orbital period, resulting 
in nearly four full Blazhko cycles.  It is clear in the figure that the Blazhko 
amplitude is changing from one Blazhko cycle to the next.  The difference between 
Blazhko maximum and minimum in the Sector 24, orbit 1 (s24o1) data is only about 5\% while four cycles 
later the change is closer to 22\%.  The shape of each Blazhko cycle is also 
changing as the amplitude grows, going from a shallow to much steeper slope 
from s24o1 to s25o2.

Figure \ref{fig:subgroups} shows the pulsation phase for the sub-group fits 
from each orbit.  The red line represents the first sub-group and the green 
line the final for each orbit.  The maximum and minimum values are plotted 
and the looping black line shows the variation in amplitude as a function 
of phase.  Comparing the graphs in Figure \ref{fig:subgroups} it is easy to see a transition in lightcurve shape from one Blazhko cycle to the next.  In particular, 
we see that the Sector 25 sub-groups are nearly sinusoidal at Blazhko minimum with 
a much steeper rise at Blazhko maximum.

Figure \ref{fig:phaseamplitude} is the corresponding phase-amplitude 
plot.  For all orbits the circulation is found to be counter-clockwise.  The 
changing shape of the phase-amplitude plot is driven primarily by the shifting 
of the maximum from the hump at $\phi$ = 0.55 to the new peak near $\phi$ = 0.75.  
This occurs very rapidly during the build-up to Blazhko maximum but much more 
slowly during the decline.  This effect can be seen in the Sector 25 data in Figure \ref{fig:subgroups}, where the blue line indicates the final sub-group before the occurrence of the peak shift.  We note that the amount of phase shift for s24o1 and s24o2 is nearly the same, although the amplitude changes are quite different.  It seems that 
the modulation of s24o1 has suppressed the peak at $\phi$ = 0.75 throughout the Blazhko 
cycle.

The overall changes seen in Figures \ref{fig:subgroups} and \ref{fig:phaseamplitude} show
characteristics that are very similar to Figures 3 and 4 from the \citet{skarka13} 
paper on the changing BE of the RRc variable, TV Boo.  Although TV Boo has much 
smaller amplitude changes than LS Her, the same type of shift in phase, due to shifting  maximum peaks, 
can be seen. TV Boo also has a very short BP of $\sim$9.7 d and an MP = 21.5 d, with possibly other, 
longer modulations that may be on the order of years. These shared characteristics suggest that LS Her 
is not an anomaly, but may represent one of a class of modulated Blazhko stars. 

\section{Comparison to Wils et al.~results}\label{sec:wils}

Due to our inability to fully measure the modulation frequencies in the 
TESS data, we decided instead to model the frequency results measured by 
\citeauthor{wils08}~to see if we could validate their results.   

To build a model using the frequencies of \citeauthor{wils08}, we analyzed 
the TESS lightcurves from s24o2 and the full Sector 25, using 
Period04. The data was pre-whitened by removing the primary frequency and 
its significant harmonics.  We then used the Blazhko and modulation triplet 
frequencies reported by Wils in Table 3 of that paper for the primary and 
second harmonics.  The Fourier series was computed and improved by fitting 
the amplitude and phase values while holding the frequencies constant.  We 
then output the resultant Fourier series to serve as our comparison model.

We excluded s24o1 in making our model in order to use it to test if the 
\citeauthor{wils08}~modulation frequencies would be able to reproduce the severely damped
Blazhko amplitude found in s24o1.

\begin{figure}[htb]
  \centering
  \begin{minipage}{0.4\textwidth}
    \centering
    \includegraphics[width=1.0\textwidth]{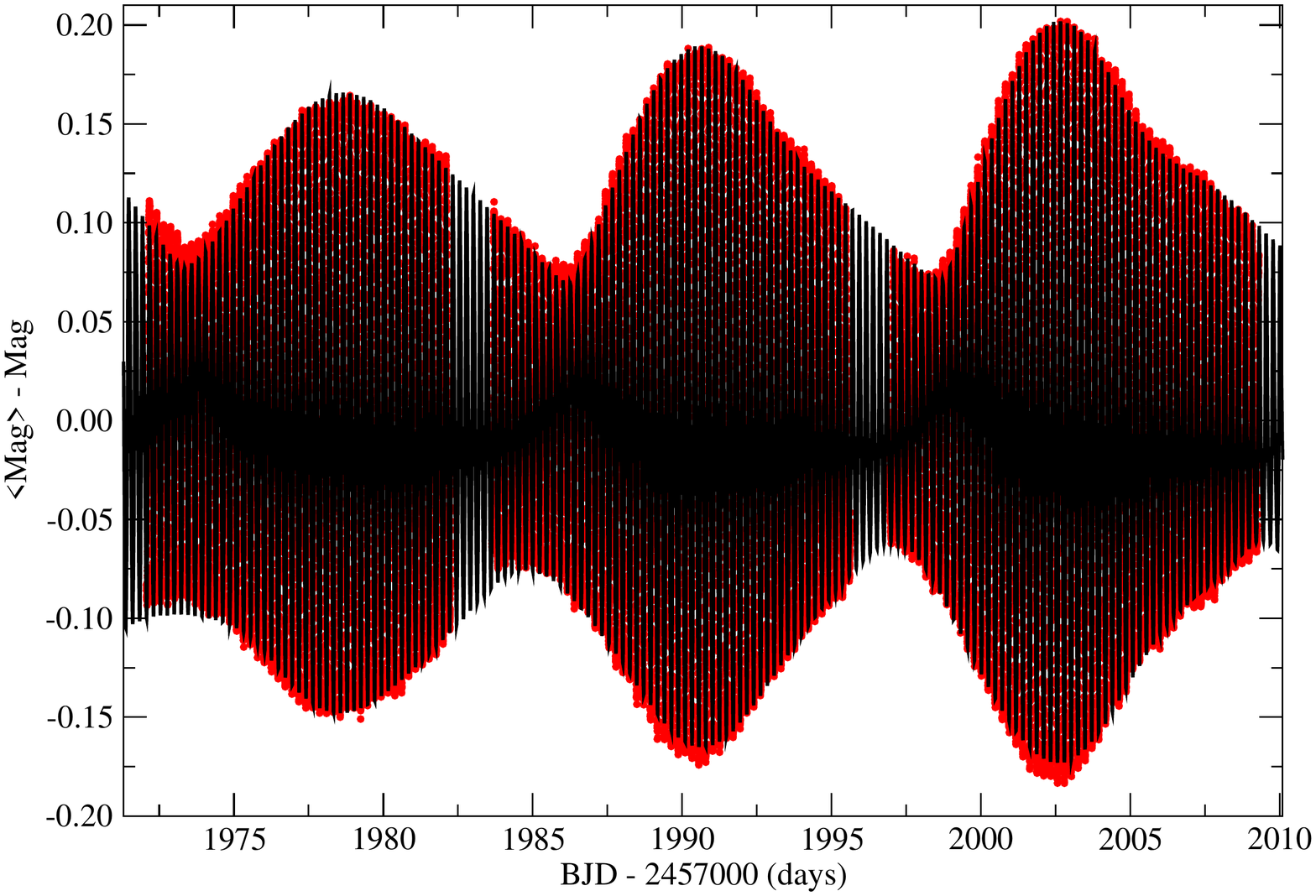}
    \label{fig:F5a}
  \end{minipage}
    \begin{minipage}{0.4\textwidth}
    \centering
    \includegraphics[width=1.0\textwidth]{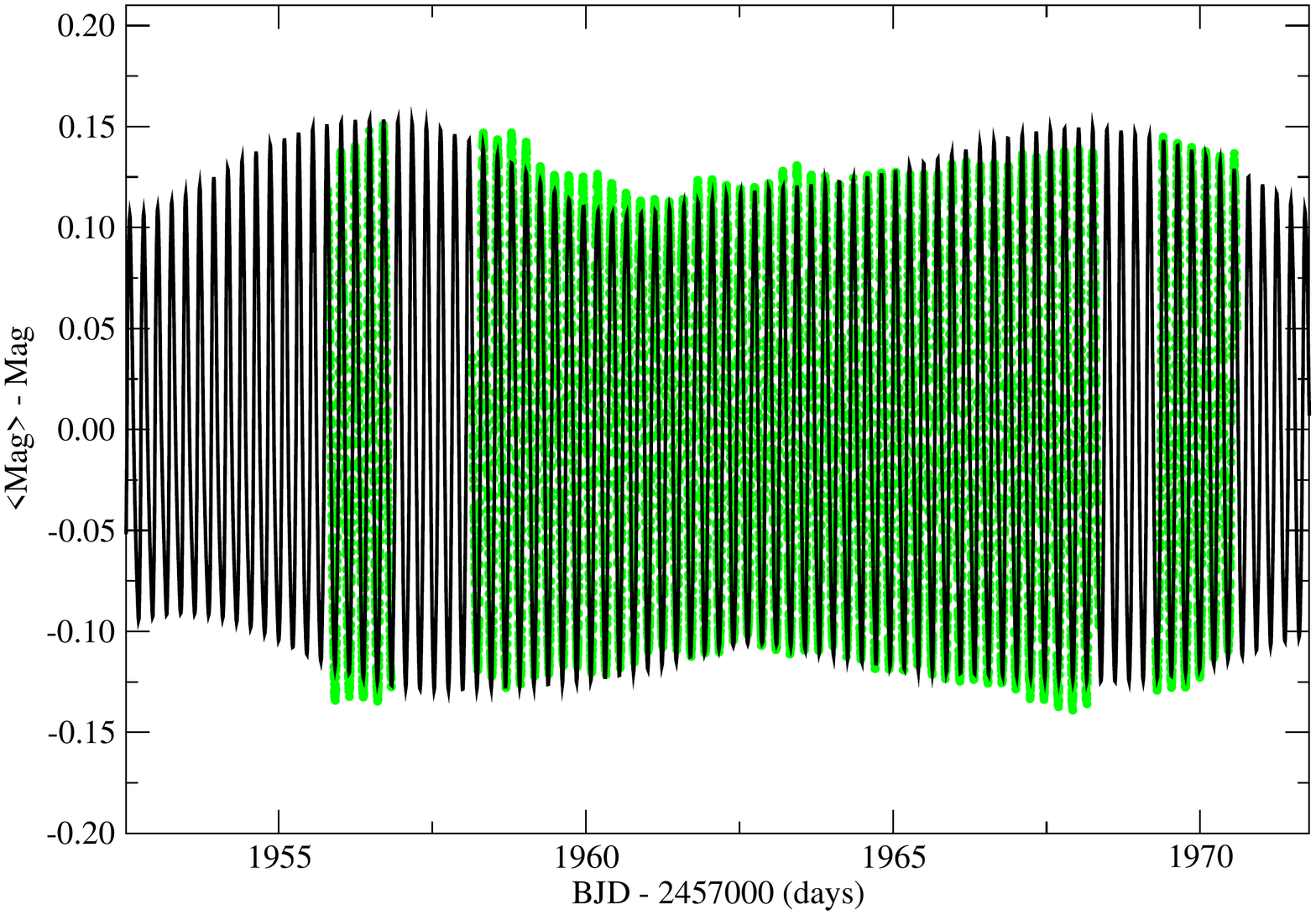}
    \label{fig:F5b}
      \end{minipage}
  \caption{The upper panel shows the model data (black) using the \citeauthor{wils08}~modulation frequencies and the TESS data (red) for s24o2, s25o1 and s25o2. This data was fit with the frequencies held constant. The lower panel shows the comparison between s24o1 and the extrapolated model data.  The s24o1 data was not used in the model fit.\label{fig:modeled}}
\end{figure}

\begin{figure}[htb]
\plotone{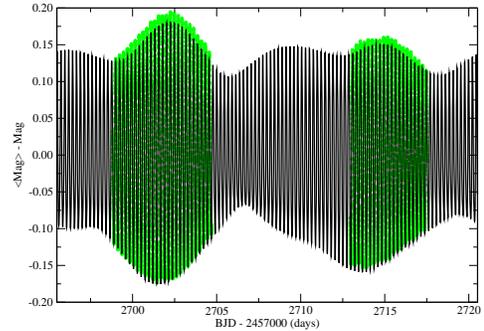}
\caption{Comparison between the model data and the TESS data from Sector 51. As in Figure 3, the TESS data is shown in green and the extrapolated model data is in black.  The comparison shows an incredibly close match.\label{fig:sec51}}
\end{figure}

The upper panel of Figure \ref{fig:modeled} shows the model fit compared to the lightcurve 
data.  Although there are a few small departures of about 2\%, overall the model is a good
representation of the data.  In the lower panel of Figure \ref{fig:modeled}, we 
compare our model data 
to s24o1, which was not fit to produce the model.  We see that the model predicts
a very damped BE, as seen in the data, again with a few small departures.  Finally, we extrapolated the 
model data to Sector 51, which was taken approximately 700 days later.  Figure \ref{fig:sec51} shows that 
the TESS data (green) is well represented by the Wils modeled data (black). The small variation at maximum light 
is only a few percent departure from the model data.  Figure \ref{fig:sec51} is in striking contrast to 
the model comparison in Figure \ref{fig:sec51misalign}.  Not only does the Blazhko maximum occur at the 
correct time, but the overall shape of the modulation matches the Sector 51 data remarkably well. We find 
this match to be strong evidence that the modulation frequencies and overall MP, as found in 
\citeauthor{wils08}, to be both accurate and stable over nearly 7 modulation cycles.

\section{Ground-based Observations}\label{sec:grou}

In July and August of 2022 we conducted additional ground based observations of LS Her in the 
Johnson-Cousins V-band.  The initial goal of these observations was to verify the damped nature 
of the Blazhko cycle as seen in s24o1.  The observations also provided further testing of the modulation effect 
by extending an additional 100 days beyond TESS Sector 51.

Four nights of observations were conducted in July 2022 at the MacAdam Student Observatory located on the 
campus of the University of Kentucky.  The data were taken with the 0.5-meter PlaneWave telescope and SBIG 
STXL-6303 research-grade CCD camera with Johnson-Cousins V and I$_\textrm{c}$ filters.  Reduction and differential 
photometry was completed using the software package AstroImageJ \citep{astroimagej}, and five reference 
stars with published V and I photometry.

A total of 18 nights of observations were taken in August 2022 using the robotic telescope of the SUTO research group\footnote{\url{www.suto.aei.polsl.pl}} located in 
Otivar, Spain (36$^\circ$48'59.99'' N 3$^\circ$40'59.99'' W). The Newtonian ASA 12'' optical tube, mounted 
on a Paramount ME, and installed in a ScopeDome 3m automatic dome, is operated remotely from the Silesian 
University of Technology (Gliwice, Poland). The observations were made using a ZWO ASI1600MM cooled CMOS 
camera (4656$\times$3520 pixels, 3.8$\mu$m pixel), which covers a field of view of 53'$\times$40' at a pixel 
scale of 0.685''. The standard image calibration, dark, and flat corrections were made using AstroImageJ. %\citep{astroimagej}. 
The custom photometric pipeline developed by the SUTO group implements classic 
aperture photometry using publicly available Python libraries, mainly 
astropy \citep{astropy} and 
PhotometryPipeline\footnote{\url{https://github.com/mommermi/photometrypipeline/blob/master/doc/index.rst}}.

To test the modulation model from the TESS data it was necessary to scale the model from the TESS magnitude 
system to that of the V-magnitudes.  We made use of the I$_\textrm{c}$ observations from MacAdam 
Observatory to set the scale between the V and I$_\textrm{c}$ measurements.  We found throughout the 
pulsation period that I$_\textrm{c}$ = 0.561$\times$V.  The 
TESS filter is broader than I$_\textrm{c}$, but is centered on this wavelength region.  We decided to 
use the above factor to scale the TESS model data.

\begin{figure}[htb]
\plotone{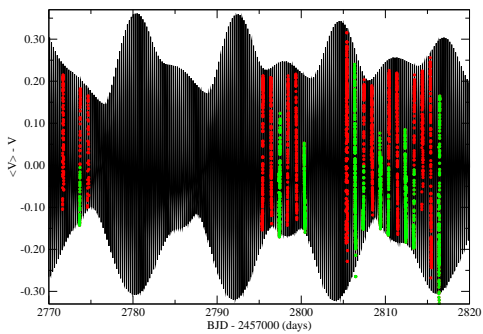}
\caption{V-magnitude observations of LS Her.  The red lightcurves include data at maximum light while the green lightcurves include data at minimum light.  The modeled data is shown in black and is extrapolated from the TESS/Wils model.  We have scaled the model data to more closely match the V-magnitude. \label{fig:vobs}}
\end{figure}

Figure \ref{fig:vobs} shows the comparison of the scaled model to the V-band 
data.  The red data shows pulsation cycles that include the pulsation maximum 
while the green includes the pulsation minimum.  The data which covered the 
minimum pulsation phase match the model data very well.  We see a few percent 
difference between the model and data maximum, which we expect is related to 
our scale factor.  In general, however, the V-band data does appear to change 
in accordance with the modulation model.  Our data did not fully cover a 
Blazhko maximum, but for the most part fell in between the maximum peaks as 
indicated by the model data. This illustrates the difficulty in establishing 
a good representation of the modulated BE from a limited ground-based data 
set.  Although we observed for a time span that is greater than the Blazhko 
period, the combination of declining Blazhko amplitude and increasing secondary 
maximum mediates the overall amplitude change. Taken alone, the result from our 
V-data shows a very limited BE that does not fully represent the range of 
Blazhko cycles.  We will discuss the possible ramifications of this in the 
following section.

\section{Discussion}\label{sec:disc}

From our analysis we can conclude that the findings of \citet{wils08} have 
been validated.  We find close agreement on the primary frequency and the 
BP for LS Her.  The MP of 109 d is also confirmed 
by our comparison to TESS and V-magnitude data, spanning 862 days.  Because 
we used the frequencies that \citeauthor{wils08}~found in their data from 
2007, we can also conclude that the modulation to the BE has very likely 
been stable over a period of 15 years.  This indicates that the modulation of 
the BE can, at least in some stars, be surprisingly stable.

\begin{figure}[htb]
\plotone{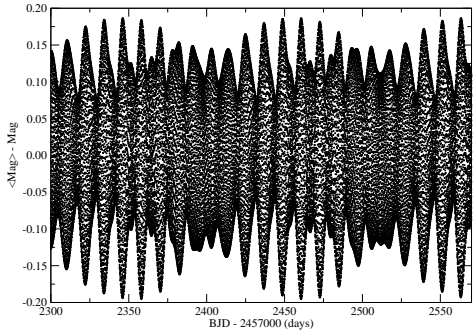}
\caption{Our modeled data shown over the course of 275 days, or 2.5 modulation cycles.  The modulated Blazhko amplitude variations can be readily seen.  Careful inspection shows the rise of a secondary hump which can destructively interfere with the Blazhko cycle, minimizing the Blazhko amplitude.\label{fig:model}}
\end{figure}

The exceptionally good representation of the model data allows us to examine 
the modulation effect in more detail.  The BE goes through large and repeated 
variations across the time span of our data.  Figure \ref{fig:model} shows 
how the modeled Blazhko cycles can look fairly normal with a growing amplitude 
for several Blazhko cycles followed by a span of $\sim$3 Blazhko cycles where 
destructive interference minimizes the BE.  During the destructive interference 
phase, the difference between the maximum Blazhko amplitude and minimum Blazhko 
amplitude is decreased by as much as 80\% of the constructive phase.  This effect 
can be readily seen in the TESS s24o1 data (Figure \ref{fig:lightcurve}) 
where the destructive phase is ending and leading into a new rising constructive 
phase.  In the V-data of Figure \ref{fig:vobs}, we see the declining BE with the 
final observations close to the destructive phase.  Depending on the time period 
of observation, it is possible to obtain greatly reduced amplitude variations, a 
more normal BE, or a confused pattern of nearly constant amplitudes mixed with 
large amplitudes.  This can particularly affect sparse, ground-based observations 
that have not fully sampled the modulation cycle.

It is known that some RRLs which have strong amplitude BEs can be observed in 
other years to be normal RRL pulsators 
\citep[e.g.,][]{jurcsik12}. If observations happen to be conducted during a destructive Blazhko modulation phase, then it would be expected that changes in amplitude will 
be minimized. This modulation effect might be difficult to detect in many Blazhko stars, 
particularly the longer period RRab stars.  LS Her has a very short BP 
of 12.75 d, and this is only about 12\% of the modulation period of 
109 d. If we scale the LS Her data to a star with a BP = 50 d, the MP is of order 400 days with the destructive phase lasting 150 days.  This 
amounts to an entire observing season over which time the BE could be greatly 
suppressed.  For data sets that may be taken years to decades apart, it becomes 
virtually impossible to detect a very long modulation effect.  We therefore 
suspect that of the sample of Blazhko stars that show Blazhko variations, at 
least some have a stable modulation like that seen in LS Her.  Currently, the short BP stars
like LS Her and TV Boo are the best probes of the modulated Blazhko effect.

As we previously noted, the MP of LS Her has likely remained 
constant for the past 15 years, spanning 440 BPs and 52 MPs.  This type of stability 
makes us wonder about its nature.  It appears 
LS Her is a fairly normal, short period, Blazhko star which undergoes stable 
modulation on a time scale of 109 d.  Although a stellar magnetic cycle has been 
previously proposed to explain the 4-year modulation of the star RR Lyrae \citep{detre73}, 
it does not seem likely that a magnetic cycle is the cause of LS Her's modulation, given
the relatively short MP and the extreme stability. 

This stability suggests that the modulation may be  driven by an external source, 
such as a binary partner. The effects of binary tidal interactions has been 
studied recently, through the analysis of ``heartbeat'' variables from the Kepler 
mission \citep{shporer}. 
These variables are caused by tidal interactions between stars on eccentric binary 
orbits.  This type of dynamic tide does not seem likely for LS Her since RRLs are 
ancient, evolved, horizontal branch stars.  It would be expected that a binary 
companion would be in a circular, possibly synchronous, orbit.  If we assume an 
equilibrium tidal effect then the companion would be expected to have an orbital 
period of P$_{\textrm{orbit}}$ = 2$\times$MP = 218 d and distance of order 1 A.U. At such 
a separation tidal effects should be inconsequential.   

In a recent paper by \citet{jurcsik18}, the authors modeled K-magnitude 
observations of a sample of Blazhko stars, to show that the variation in 
amplitude during the Blazhko cycle is due primarily to changes in the changing 
surface temperature of the stars, rather than the effect of variations in the 
changing radius.  In the case of LS Her, we might expect that temperature 
variations will be observed during a Blazhko cycle but that variations in the 
changing radius are responsible for the modulation effect. Given that LS Her is 
a relatively bright variable with $\langle$V$\rangle$ = 10.8, it is a prime candidate for 
follow-up, high-resolution spectroscopic observations. We intend to obtain 
spectral observations of LS Her in the coming year.  These observations will 
provide the kinematics to test for binarity and to test for variations in the 
changing radius as a function of Blazhko and modulation cycle.  

Finally, the first-overtone, RRc, Blazhko variables seem to be, in general, 
good probes of the modulated BE.  The short BPs and 
MPs, compared to the RRab Blazhko stars, allows a much shorter 
time span to constrain the modulation effect. Future detailed studies should 
include the RRc variables as test beds of the Blazhko effect. 

\begin{acknowledgments}
Members of SUTO team were responsible for automation and running remote 
observations at Otivar observatory and for data processing.\\
R.W.~would like to acknowledge funding from the National Science Foundation, 
REU program (\#1950795) at University of Kentucky.  The work of H.H.M. was 
also funded through this program.\\
K.C.~would like to acknowledge funding from the National Science Foundation 
LEAPS-MPS program through Award \#2137787. The work of J.L., J.T., and 
S.Y.~was also supported through this award.\\
\end{acknowledgments}

\bibliography{biblio}{}
\bibliographystyle{aasjournal}

\end{document}